\documentclass{llncs}
\usepackage{makeidx}  

\usepackage{amsmath}
\usepackage{amsfonts,amssymb}
\usepackage{clrscode3e}

\begin{document}
\frontmatter          
\pagestyle{headings}  
\addtocmark{Graph isomorphism} 
\title{Reduction of the graph isomorphism problem\\
to equality checking of $n$-variable polynomials\\
and the algorithms that use the reduction}
\titlerunning{Graph isomorphism}  
%
\author{Alexander Prolubnikov}
\authorrunning{Alexander Prolubnikov} 
%
\tocauthor{Alexander Prolubnikov}
\institute{Omsk State University, Omsk, Russian Federation\\
\email{a.v.prolubnikov@mail.ru}
}
\maketitle              

\begin{abstract}

The graph isomorphism problem is considered. We assign modified $n$-variable characteristic polynomials for graphs and reduce the graph isomorphism problem to the problem of the polynomials isomorphism. It is required to find out, is there such a numbering of the second graph's vertices that the polynomials of the graphs are equal. 

\smallskip

We present algorithms for the graph isomorphism problem that use the reduction. We prove the propositions that justify the possibility of numerical realization of the algorithms for the general case of the graph isomorphism problem. The algorithms perform equality checking of graphs polynomials. We show that pro\-ba\-bility of obtaining a wrong solution of the graph isomorphism problem by comparing values of graph polynomials is negligible if the mantissa length is sufficiently large. 

\smallskip

Since, for a graph on $n$ vertices, the graph polynomial has $2^n$ coefficients, its value at some point cannot be evaluated directly for large enough $n$. We show that we can check the equality of the polynomials at some points without direct evaluation of the polynomials values at these points. We prove that it is required $O(n^4)$ elementary machine operations and machine numbers with mantissas length $O(n^2)$ to check equality of the values  for the graphs on $n$ vertices. 

\smallskip

For the worst, it needs an exponential from $n$ time to solve the graph isomorphism problem instance using the presented approach, but in practice, it is efficient even for well known computationally hard instances of the graph isomorphism problem.

\end{abstract}
\section{The graph isomorphism problem}
In the graph isomorphism problem {\it (GI)}, we have two simple graphs $G$ and $H$. 
Let $V(G)$ and $V(H)$ denote the sets of vertices of the graphs and let $E(G)$ 
and $E(H)$ denote the sets of their edges. $V(G)\!=\!V(H)\!=\![n]$. An 
isomorphism of the graphs $G$ and $H$ is a bijection $\varphi:V(G)\to V(H)$ 
such that for all $i,j\!\in\!V(G)$ 
$$(i,j)\in E(G)\Leftrightarrow (\varphi(i),\varphi(j))\in E(H).$$ If such 
a bijection exists, then the graphs $G$ and $H$ are isomorphic (we denote it as $G\!\simeq\! H$), else the graphs are not isomorphic. To solve GI, 
it is required to present the bijection that is an isomorphism or we must show 
non-existence of such a bijection.

We may formulate GI in terms of adjacency matrices of graphs. Let $(A)_{ij}$ 
denote the $ij$-th element of a matrix $A$. The adjacency matrix of a graph $G$ 
is a matrix $A(G)$ which has dimension of $n\!\times\! n$. Elements of this matrix 
are defined as follows: 
$$(A(G))_{ij}=\left\{
\begin{array}{ll}
1,&\mbox{ if } (i,j)\in E(G),\\
0,&\mbox{ else}.\\
\end{array}
\right.$$ Let $A(G)$ and $A(H)$ be adjacency matrices of the graphs $G$ and $H$. 
Then $$
G\simeq H\Leftrightarrow \exists \varphi\in S_n 
:\ A(H)=P_{\varphi}A(G)P_{\varphi}^\top,
$$ where $S_n$ is a symmetric group on $[n]$ and
$$
(P_{\varphi})_{ij}=\left\{
\begin{array}{ll}
1,&\mbox{ if } i=\varphi(j),\\
0,&\mbox{ else}.\\
\end{array}
\right.$$ 
By this formulation of the problem, the two graphs are isomorphic if and only if 
we can obtain adjacency matrix of one of them from adjacency matrix of the 
second by some permutation of its rows and columns.

GI is one of the fundamental problems of discrete mathematics and there are 
numerous applications where it arises. For example, without 
solving GI, we cannot practically solve the problem which may be formulated as searching of $n$-vertices graph that has some specified property. It is may be structure graph of a synthesized chemical compound. We can search the graph doing the exhaustive search on all labelled $n$-vertices graphs. But, since there are $n!$ of isomorphic labelled graphs for every unlabeled graph that identify the compound, we must check the property only for nonisomorphic labeled graphs during this search. For this purpose, we need to efficiently solve GI. Else our time and memory expenses would be too high. 

GI belongs to class NP since it takes $O(n^2)$ time to check whether some 
bijection of $V(G)$ onto $V(H)$ is an isomorphism. It has not been proven that the problem 
is NP-complete and there is no polynomial algorithm has been 
obtained for the general case of the problem.

GI is solvable in polynomial time for some classes of graphs: for trees 
\cite{Lindell}, for graphs with bounded genus \cite{Filotti}, for graphs with 
bounded multiplicity of their adjacency matrix eigenvalues \cite{Babai}, for 
graphs with bounded degree of their vertices \cite{Luks} and for some other 
restricted classes of graphs. The more regular structure of the graphs, the harder 
to obtain solution of GI for them. Such classes as strongly regular graphs, isoregular graphs give the instances of GI that cannot be solved in polynomial time by existing algorithms.

For GI, the designed algorithms may be divided in two classes.
The first class algorithms are designed to solve GI for some restricted cases 
and the second class algorithms solve the problem for the general 
case. The examples of the algorithms which belongs to the first class are the 
algorithms that solve GI for the mentioned above classes of graphs. The Ullman 
algorithm \cite{Ullman}, the Schmidt-Druffel algorithm \cite{Schmidt}, B. McKay's 
NAUTY algorithm \cite{McKay} belong to the second class of the algorithms. These 
algorithms are exponential in the worst case.

To solve GI, the algorithms check graph invariants during 
their implementation. {\it Graph invariants} are properties of graphs which 
are invariant under graph isomorphisms, i.e., graph invariant is a function 
$f$ such that $f(G)\!=\!f(H)$ if $G\!\simeq\! H$. The examples 
of graph invariants are such graph properties as connectivity, genus, 
degree sequence, the characteristic polynomial of adjacency matrix, its 
spectrum. A graph invariant $f(G)$ is called {\it complete} if  
the equality $f(G)=f(H)$ implies that $G\!\simeq\! H$. 

Let us consider some of well known graph invariants. Applying the Weisfeiler-Lehman method, we perform 
iterative classification of graphs vertices based on distances between them. As a result,
we have such colourings of vertices that we may use it to distinguish 
non-isomorphic graphs. Using this method, we can solve GI for large 
class of graphs in polynomial time. But it is shown \cite{Cai} that there 
exists such pairs of non-isomorphic graphs on $n$ vertices that they are 
cannot be distinguished by $k$-dimensional Weisfeiler-Lehman algorithm in 
polynomial time for $k\!=\!\Omega(n)$. This implies that, for the general case, 
this method not solve GI in polynomial time.

The procedures of a {\it graph canonization} give complete graph invariants. For a 
graph $G$, using some procedure of graph canonization, we obtain its {\it canonical 
form} that is some labeled graph assigned to $G$. Two graphs are isomorphic if and only if they 
have the same canonical form. Using canonical form of a graph, we may compute 
its canonical code. {\it Canonical code} $c(G)$ of a graph $G$ is a bit string (or it can be represented as a bit string) such that $G\!\simeq\! H$ if and only if $c(G)\!=\!c(H)$. The example of a canonical 
code is the code $c_0(G)$: $$c_0(G)=\max\limits_{\pi\in S_n}\{(A)_{\pi(1)}||(A)_{\pi(2)}||\ldots||(A)_{\pi(n)}\},$$
where ``$||$'' denotes concatenation of bit strings and $(A)_i$ denotes the 
$i$-th row of the adjacency matrix $A(G)$. I.e., $c_0(G)$ is the maximum number that we can get concatenating permutated rows of $A(G)$. For some classes of graphs, canonical 
codes may be computed in polynomial time \cite{Lindell}, \cite{Datta}. 
Every complete invariant gives a way for graph canonization.

In \cite{Grigoriev}, a complete invariant for hypergraphs is presented. 
This complete invariant is not a result of canonization. 
For the case of simple graphs, this invariant is a system of $n^2+1$ 
polynomials over a field of characteristic $q$, where $q$ is a prime number or 
zero. 

\smallskip

In our work, we assign modified characteristic polynomials for graphs 
and reduce the graph isomorphism problem to the problem of the polynomials isomorphism. To solve this problem, it is
required to find out, is there such a numbering of the second polynomial's variables  
that the polynomials are equal. 

Since the graph polynomials we consider have $2^n$ coefficients, we cannot check equality of such polynomials in polynomial time. But we show that we can check the equality of the 
polynomials' values at some points without direct computation of the values. 
We prove that, for the graphs on $n$ vertices, it is required $O(n^4)$ 
elementary operations and it is required machine numbers with mantissa 
length $O(n^2)$ to numerically check the equality of the graphs polynomials 
values at some point of $\mathbb{R}^n$.                             

For the worst, it needs an exponential from $n$ time to solve the graph isomorphism problem instance using the presented approach, but in practice, it is efficient even for well known compuationally hard instances of the graph isomorphism problem.

\section{The modified characteristic polynomial of a graph}

\subsection{Characteristic polynomial of a graph and its modifications} 

Let us consider the characteristic polynomial of a graph and some of its 
modifications which have been used for characterization of graphs by their 
structural properties. The cha\-racteristic polynomial of a graph $G$ is the 
polynomial $$\chi_{\mbox{\tiny{$G$}}}(x)=\det(A(G)-xE),$$ where $x$ is a 
variable and $E$ is the identity matrix. Let $d_i$ denote {\it degree} of the vertex $i\!\in\!V(G)$, i.e., it is the number of edges incident to the vertex. Let $D(G)=\mbox{diag}(d_1,\ldots,d_n)$. 

Some modifications of the characteristic polynomial 
are considered in \cite{Cvetkovic}. The examples are the graph Laplacian $L(G)$ 
that is defined as $$L(G)=D(G)-A(G),$$ the signless graph Laplacian $|L(G)|$ 
that is defined as $$|L(G)|=D(G)+A(G),$$ and some other 
modifications which may be generalized by the polynomial $\xi_G(x,y)$: $$\xi_{\mbox{\tiny{$G$}}}(x,y)=\det (xE-(A(G)-yD(G))).$$ 

Seidel polynomial \cite{Seidel} is another modification of the characteristic 
polynomial that is obtained by modification of a graph adjacency matrix. It is 
the polynomial $\zeta_{\mbox{\tiny{$G$}}}(x)$: $$\zeta_{\mbox{\tiny{$G$}}}(x)=\det(xE-(F-E-2A(G))),$$ 
where $(F)_{ij}=1$ for $i,j=\overline{1,n}$. 

A generalization of the characteristic polynomial is the polynomial that is 
presented in \cite{Lipton}. It is a polynomial $\psi_{\mbox{\tiny{$G$}}}(x,y,\lambda)$ 
of the form $$\psi_{\mbox{\tiny{$G$}}}(x,y,\lambda)=\det(A(x,y)-\lambda E),$$ 
where $A(x,y)$ is the matrix derived from $A$ in which $1$s are replaced by 
variable $x$ and $0$s (other than the diagonals) are replaced by variable $y$.

None of the presented modifications of the characteristic polynomial is a 
complete graph invariant. I.e., there exist non-isomorphic graphs with the same polynomials for the mentioned types of polynomials.

\subsection{The modified characteristic polynomial of a graph} 

Variables of 
the polynomials presented above have no connection with graph vertices. We 
modify the characteristic polynomial $\chi_G(x)$ of a graph $G$ on $n$ vertices 
assigning the variable $x_i$ to the vertex $i\!\in\!V(G)$. Let $x_1,\ldots,x_n$ be 
independent variables and let $X=\mbox{diag}(x_1,\ldots,x_n)$. 

\smallskip

\begin{definition}
For a graph $G$, $|V(G)|=n$, $\eta_{\mbox{\tiny{$G$}}}(x_1,\ldots,x_n)$ 
is a polynomial of the form 
\begin{equation}\label{1.1}
\eta_{\mbox{\tiny{$G$}}}(x_1,\ldots,x_n)=\det(A(G)+X).
\end{equation} 
\end{definition}

\bigskip

For graphs on $n=1,2,3$ vertices, the polynomials of the form (\ref{1.1}) 
are the following ones:

1) $n=1$: $x_1$;

2) $n=2$: $x_1x_2$, $x_1x_2\!-\!1$;

3) $n=3$: $x_1x_2x_3$, $x_1x_2x_3\!-\!x_1$, $x_1x_2x_3\!-\!x_1\!-\!x_3$, $x_1x_2x_3\!-\!x_1\!-\!x_2\!-\!x_3\!+\!2$.

\noindent It is clear that we have different polynomials for 
non-isomorphic graphs no matter what numbering of its vertices we use 
for $n\!=\!1,2,3$.

For a subset $c$ of $V(G)$, let $x_c$ be a product of the form 
$\prod_{i\in c} x_i$ and let $A(G)_c$ be the determinant of the submatrix of $A(G)$ 
that is obtained by deleting of rows and columns of $A(G)$ which numbers 
belong to the subset $c$. $A(G)_c$ is a coefficient of $\prod_{i\in c} x_i$ in the
polynomial $\eta_{\mbox{\tiny{$G$}}}(x_1,\ldots,x_n)$. Having $c$ and $\varphi\!\in\! S_n$, let 
$\varphi(c)$ be the image of $c$: $\varphi(c)\!=\!\{\varphi(i)\ |\ i\!\in\! c\}$. 
For $x\!=\!(x_1,\ldots,x_n)$, $x_{\varphi}\!=\!(x_{\varphi(1)},\ldots,x_{\varphi(n)})$. 
The following theorem holds.

\bigskip

\begin{theorem} $G\!\simeq\! H$ and $\varphi\! :\! V(G)\!\to\! V(H)$ is an isomorphism of the graphs if and only if, for all $x\!\in\!\mathbb{R}^n$,  
\begin{equation}\label{1.2}
\eta_{\mbox{\tiny{$G$}}}(x_1,\ldots,x_n)=\eta_{\mbox{\tiny{$H$}}}(x_{\varphi(1)},\ldots,x_{\varphi(n)}).
\end{equation}
\end{theorem}

The polynomials $\eta_{\mbox{\tiny{$G$}}}$ and $\eta_{\mbox{\tiny{$H$}}}$ are {\it isomorphic} if they are equal for some numbering of variables of $\eta_{\mbox{\tiny{$H$}}}$. The Theorem 1 states that two graphs are isomorphic if and only if their polynomials of the form (\ref{1.1}) are isomorphic. There is no such numbering that gives equal polynomials for non-isomorphic graphs. 

The polynomials are equal if the coefficient of $x_c$ is equal to the coefficient of $x_{\varphi(c)}$ for every subset $c$ of $V(G)$. Thus, the equality (\ref{1.2}) holds if and only if $A(G)_c\!=\! A(H)_{\varphi(c)}$ for all subsets $c$ of $V(G)$. 

\smallskip

Let us prove the Theorem 1.

\begin{proof} If $G\!\simeq\! H$ and $\varphi$ is an isomorphism of the graphs, then 
(\ref{1.2}) holds since 
\begin{equation}\label{1.3}
A(H)=P_{\varphi}A(G)P_{\varphi}^\top,
\end{equation}
and the coefficients of $\eta_{\mbox{\tiny{$G$}}}$ and $\eta_{\mbox{\tiny{$H$}}}$ 
corresponded by $\varphi$ are equal to each other.

Let us show that if the equality (\ref{1.2}) holds, then $G\!\simeq\! H$ and $\varphi$ 
is an isomorphism of the graphs. Let us denote $A(G)$ as $A\!=\!(a_{ij})$ and $A(H)$ as 
$B\!=\!(b_{ij})$.

If (\ref{1.2}) holds for all $x\!\in\!\mathbb{R}^n$, then the coefficient of $x_c$ is equal to the 
coefficient of $x_{\varphi(c)}$ for any subset $c$ of $V(G)$.
Thus, if we take $c$ such that $c\!=\!V(G)\!\setminus\!\{i,j\}$ for some pair of vertices 
$i,j\!\in\! V(G)$, we have $A_c\!=\!B_{\varphi(c)}$. This is equivalent to $
\det 
\begin{pmatrix}
0 & a_{ij} \\
a_{ij} & 0 
\end{pmatrix} 
\!=\!
\det 
\begin{pmatrix}
0 & b_{\varphi(i)\varphi(j)} \\
b_{\varphi(i)\varphi(j)} & 0 
\end{pmatrix}. 
$
So $a_{ij}\!=\!b_{\varphi(i)\varphi(j)}$, and $(i,j)\!\in\! E(G)$ if and only if $(\varphi(i),\varphi(j))\!\in\! V(H)$,
i.e., $G\!\simeq\! H$ and $\varphi$ is an isomorphism.\qed 
\end{proof}

\noindent{\bf Remark.} To check the equality (\ref{1.2}) for some $\varphi$, 
it is sufficient to check equality of only coefficients that correspond to $c$ such that 
$|c|=n\!-\!2$. Since if $A_c=B_{\varphi(c)}$ for such $c$, then $A(H)$ may be 
obtained from $A(G)$ by permutation of its rows and columns. So $A_c=B_{\varphi(c)}$ for all subsets $c$ of 
$V(G)$. 

\section{Equality checking of the modified characteristic\\ polynomials by checking 
equality of their coefficients}

For $I\!=\!\{i_1,\ldots,i_{m}\}\!\subseteq\! V(G)$, let $C_I^k$ be 
a set of all subsets with $k$ elements of $I$, let $C_I=\cup_{k=1}^{|I|} C_I^k$. 
For $c\!\in\! C_I$, $e_c\!=\!\sum_{i\in c}e_i$, $\varepsilon_c=\varepsilon\cdot e_c$, 
where $\varepsilon\!\in\!\mathbb{R}$, $\varepsilon>0$. $\{e_i\}_{i=1}^n$ is a 
standard basis of $\mathbb{R}^n$.

The presented below $\proc{Algorithm $1$}$ compares coefficients of  
polynomials of the form (\ref{1.1}) using the recursive procedure 
$\proc{Equality checking of coeffici-}$ $\proc{ents}$. This procedure checks  
equality of the coefficients using the fact that the polynomials 
are linear in every variable (they are {\it polylinear} functions). In the course of implementation of the algorithm, we trying to set the correspondence (bijection) $\varphi$. 
Set $I$ contains such vertices of $V(G)$ that the correspondence $\varphi$ 
is established for them after an iteration is performed. $J\!=\!\varphi(I)$. If the variable $a$ takes the 
value $b$, we denote it as $a\!\leftarrow\! b$. 

\begin{codebox}
\Procname{
$\proc{Equality checking of coefficients}\ (i)$}
\li \If $i\!=\!n$ 
\li \Then $flag\!\leftarrow\! true$;
\li {\bf exit} 
\li \Else 
\li \For $j\!\leftarrow\! 1$ \To $n$ \Do 
\li \If ($j\!\not\in\! J$ and $\forall c\!\in\! C_I :\ 
\eta_{\mbox{\tiny{$G$}}}(\varepsilon_c+\varepsilon e_i)\!=\!\eta_{\mbox{\tiny{$H$}}}(\varepsilon_{\varphi(c)}+\varepsilon e_j)$) 
\li \Then $\varphi(i)\!\leftarrow\! j$;
\li $I\leftarrow I\!\cup\!\{i\}$; 
\li $J\leftarrow J\!\cup\!\{j\}$;
\li $\proc{Equality checking of coefficients}\ (i+1)$;
\End
\End
\End
\li $I\leftarrow I\!\setminus\!\{i\}$;
\li $J\leftarrow J\!\setminus\!\{\varphi(i)\}$;
\li the value $\varphi(i)$ is not defined;
\li $flag\leftarrow false$. 
\end{codebox}

\begin{codebox}
\Procname{
$\proc{Algorithm 1}\ (G, H)$}
\li $I\!\leftarrow\!\varnothing$; $J\!\leftarrow\!\varnothing$; the values $\varphi(i)$ 
are not defined for $i\!=\!\overline{1,n}$; 
\li $\proc{Equality checking of coefficients}\ (1)$;
\li \If $flag$ 
\li \Then {\bf print} ``$G\!\simeq\! H$, $\varphi$ is an isomorphism 
of $G$ and $H$'';
\li \Else 
\li {\bf print} ``$G\!\not\simeq\! H$''.
\End
\end{codebox}

\noindent Implementing $\proc{Equality checking...}$ 
procedure, we search for $j\!\in\!V(H)\!\setminus\! J$ such that
\begin{equation}\label{2.1}
\eta_{\mbox{\tiny{$G$}}}(\varepsilon_c+\varepsilon e_i)=
\eta_{\mbox{\tiny{$H$}}}(\varepsilon_{\varphi(c)}+\varepsilon e_j).
\end{equation} We check this equality for every $c\!\in\! C_I$. If there is 
no such $j$, then we exit from the procedure 
having $flag\!=\!false$. We have $flag\!=\!true$ if and only if we have 
set an isomorphism $\varphi$ of the graphs $G$ and $H$.

For $c\!=\!\{i_1,\ldots,i_k\}$, the coefficient $A_c$ is written below as $A_{\{i_1,\ldots,i_k\}}$.

At the start of the algorithm, we have $I\!=\!\varnothing$, $C_I\!=\!\varnothing$, 
$J\!=\!\varnothing$. For $i\!=\!1$, checking of the equality (\ref{2.1}) is 
eqiuvalent to checking of the equality
\begin{equation}\label{2.2}
\eta_{\mbox{\tiny{$G$}}}(\varepsilon e_1)=\eta_{\mbox{\tiny{$H$}}}(\varepsilon e_j)
\end{equation} for $j\!\in\!V(H)$.
(\ref{2.2}) is equivalent to
\begin{equation}\label{2.3}
\det A+A_{\{1\}}\varepsilon =\det B+B_{\{j\}}\varepsilon.
\end{equation} 
If (\ref{2.3}) holds for $\varepsilon\!=\!0$ and for some $\varepsilon\!>\!0$, 
then $\det A\!=\!\det B$, $A_{\{1\}}\!=\!B_{\{j\}}$. If it is so, 
we set $\varphi(1)\!\leftarrow\! j$. Further, if it occurs 
that we exit from the procedure $\proc{Equality checking...}$ with 
$flag\!=\!false$, then the value of $\varphi(1)$ may become undefined again.

For $i\!=\!2$, when we have $I\!=\!\{1\}, C_I\!=\!C_I^1=\!\{\{1\}\}$, 
$J\!=\!\{\varphi(1)\}$, we need to find $j\!\in\!V(H)\!\setminus\!J$\ \ such that
\begin{equation}\label{2.4}
\eta_{\mbox{\tiny{$G$}}}(\varepsilon e_2)=\eta_{\mbox{\tiny{$H$}}}(\varepsilon e_j)
\end{equation}
and
\begin{equation}\label{2.5}
\eta_{\mbox{\tiny{$G$}}}(\varepsilon e_1+\varepsilon e_2)=
\eta_{\mbox{\tiny{$H$}}}(\varepsilon e_{\varphi(1)}+\varepsilon e_j).
\end{equation}
Checking of the equality (\ref{2.4}) is equivalent to checking of the equality
\begin{equation}\label{2.6}
\det A+A_{\{2\}}\varepsilon=\det B+B_{\{j\}}\varepsilon,
\end{equation} 
and checking of the equality (\ref{2.5}) is equivalent to checking of the equality
\begin{equation}\label{2.7}
\det A+A_{\{1\}}\varepsilon+A_{\{2\}}\varepsilon+A_{\{1,2\}}\varepsilon^2=
\det B+B_{\{\varphi(1)\}}\varepsilon+B_{\{j\}}\varepsilon+B_{\{\varphi(1),j\}}
\varepsilon^2.
\end{equation}
If (\ref{2.3}) holds for $\varepsilon\!=\!0$ and for some $\varepsilon\!>\!0$, then, 
if (\ref{2.6}) holds for some $\varepsilon\!>\!0$, we have $A_{\{2\}}\!=\!B_{\{j\}}$, 
and, if (\ref{2.7}) holds too, we have $A_{\{1,2\}}\!=\!B_{\{\varphi(1),j\}}$. 
If it is so, we set $\varphi(2)\!\leftarrow\! j$.
Further, if it occurs that we exit from the procedure 
$\proc{Equality checking...}$ with $flag\!=\!false$, 
then the value of $\varphi(2)$ may become undefined again.

Thus, at the moment when we check equality of $A_{c\cup\{i\}}$ and 
$B_{\varphi(c)\cup\{j\}}$, we have $A_c\!=\!B_{\varphi(c)}$ 
for all $c\!\in\!C_I$ since otherwise there 
would be exit from the procedure $\proc{Equality checking...}$ 
with $flag\!=\!false$. 
So
\begin{equation}\label{2.8}
\eta_{\mbox{\tiny{$G$}}}(\varepsilon_c+\varepsilon e_i)=
\sum\limits_{c'\in P(c)}\varepsilon^{|c'|}A_{c'}
+\varepsilon^kA_{c\cup\{i\}},
\end{equation}
\begin{equation}\label{2.9}
\eta_{\mbox{\tiny{$H$}}}(\varepsilon_{\varphi(c)}+\varepsilon e_j)=\sum\limits_{\varphi(c')\in Q(c)}\varepsilon^{|c'|}B_{\varphi(c')}
+\varepsilon^kB_{\varphi(c)\cup\{j\}},
\end{equation}
where $P(c)\!=\!\{c'\!\in\! C_{I\cup\{i\}}\, |\, c'\!\subset\! c\}$, 
$Q(c)\!=\!\{\varphi(c')\!\in\! C_{\varphi(I)\!\cup\!\{j\}}\, |\, c'\!\subset\! c\}$,
$k\!=\!|c|\!+\!1\!>\!|c'|$ for all $c'\!\in\!P(c)$. Since, 
for all $c'\!\in\!P(c)$, we have $A_{c'}\!=\!B_{\varphi(c')}$ at the moment when 
we check the equality (\ref{2.1}) for $\varphi(c')\!\in\!Q(c)$, then, if (\ref{2.1}) 
holds for $c\!\in\! C_I$, it follows by (\ref{2.8}) and (\ref{2.9}) that we 
have $A_{c\cup\{i\}}\!=\!B_{\varphi(c)\cup\{j\}}$. Setting 
$\varphi(i)\!\leftarrow\! j$ and $I\!\leftarrow\! I\cup\{i\}$, $J\!\leftarrow\! J\cup\{j\}$, 
we obtain $A_c\!=\!B_{\varphi(c)}$ for all $c\!\in\! C_I$. 

As a result, if we exit from the procedure 
$\proc{Equality checking...}$ with $flag\!=\!true$,
then we have find a bijection $\varphi$ such that $A_c=B_{\varphi(c)}$ for all 
$c\!\in\! C_I$, where $I=[n]$. Thus, by Theorem 1, $G\!\simeq\! H$ 
and $\varphi$ is an isomorphism of the graphs. It proves the Proposition 1.

\begin{proposition} 
GI can be solved by comparison of coefficients of graphs polynomials for which we perform comparison of their values at least at $2^n$ points in~$\mathbb{R}^n$.
\end{proposition}

\section{Solving GI by checking equality of the graph polynomials values at predefined points}

In the procedure above, we do a comparison of two polynomials with renumbered variables for the second one at least at $2^n$ points, which is necessary to check equality of polynomials with $2^n$ coefficients. Such algorithm has an exponential complexity no matter what instance of GI it solves. The algorithms presented below require finding no more than $n$ points in $\mathbb{R}^n$ to implement checking equality of graph polynomials with a negligible probability of mistake. This approach make possible to significantly reduce the time that is needed to solve GI instance. However, finding these points itself may require exponential time.

\subsection{The Direct algorithm for GI}

Let $N\!\in\!\mathbb{N}$, $S\!=\!\{k/10^N\ |\ 0\!<\!k\!<\! 10^N,\ k\!\in\!\mathbb{Z}_+\}$, 
$S\!\subset\!(0,1)$. For $i\!=\!\overline{1,n}$, let $\varepsilon_i$ 
be selected at random independently and uniformly from $S$. 
Let $\varepsilon^{(i)}$ be the following points of $\mathbb{R}^n$:
$\varepsilon^{(0)}\!:=\!0$, $\varepsilon^{(i)}\!:=\!\varepsilon^{(i-1)}\!+\!\varepsilon_ie_i$, $i\!=\!\overline{1,n}$. I.e., $\varepsilon^{(1)}\!=\!(\varepsilon_1,0,\ldots,0)$, 
$\varepsilon^{(2)}\!=\!(\varepsilon_1,\varepsilon_2,\ldots,0)$,
$\ldots$, $\varepsilon^{(n)}\!=\!(\varepsilon_1,\varepsilon_2,\ldots,\varepsilon_n)$, $\varepsilon_i\!\neq\!\varepsilon_j$.

In the course of implemetation of the algorithms we present below, we trying 
to set a bijection $\varphi\!:\!V(G)\!\to\! V(H)$ such that
$\eta_{\mbox{\tiny{$G$}}}(\varepsilon^{(i)})\!=\!\eta_{\mbox{\tiny{$H$}}}(\varepsilon_{\varphi}^{(i)})$.
For $i\in V(G)$, we searching for such $j\!\in\! V(H)$ that 
\begin{equation}\label{3.1}
\eta_{\mbox{\tiny{$G$}}}(\varepsilon^{(i)})=
\eta_{\mbox{\tiny{$H$}}}(\varepsilon_{\varphi}^{(i-1)}+\varepsilon_i e_j).
\end{equation}
If we have set up such $\varphi$ for the graphs $G$ and $H$ in the course of $n$ consecutive iterations of the algorithm, then we make a conclusion that the graphs 
are isomorphic and $\varphi$ is an isomorphism, else we make 
a conclusion that they are not isomorphic. 

The Direct algorithm for GI is titled below as 
$\proc{Algorithm $2$}$. It is a test for isomorphism that may be mistaken 
for some instances of GI. For any test for isomorphism, there are two kinds 
of mistakes it can make: 1) a wrong conclusion that $G\!\simeq\!H$, when 
$G\!\not\simeq\! H$, 2) a wrong conclusion that $G\!\not\simeq\! H$, when 
$G\!\simeq\! H$. As it shall be stated below, the probability of a 
mistake of the first kind can be considered as negligible for the Direct 
algorithm while there may be mistake of the second kind. 

The Direct algorithm solves the GI instances that presented in 
\cite{Foggia} but it makes mistake of the second kind for GI 
instances obtained for strongly-regular graphs from \cite{SpenceLib} 
when $n\!\ge\!13$.
 
\subsection{Recursive modification of the Direct algorithm}

The $\proc{Algorithm $3$}$ is a recursive modification of the 
Direct algorithm. It~searches for such points $\varepsilon^{(i)}$ and $\varepsilon_{\varphi}^{(i)}$ that 
$\eta_{\mbox{\tiny{$G$}}}(\varepsilon^{(i)})=\eta_{\mbox{\tiny{$H$}}}(\varepsilon_{\varphi}^{(i)})$, $i=\overline{1,n}$, using the procedure $\proc{Set the correspondence}$.

The recursive procedure $\proc{Set the correspondence}$ gets 
on input $i\!\in\!V(G)$ and set $\varphi(i)\!\leftarrow\! j$ 
for $j\!\in\!J\!\subseteq\! V(H)$ such that (\ref{3.1}) holds. If there 
is no such $j\!\in\!J$, then we modify the correspondence that was already setted 
up for $i\!-\!1$: we use the next element of $J$\ \ for setting 
$\varphi(i\!-\!1)$.

In addition to the GI instances that presented in \cite{Foggia}, the $\proc{Algorithm $3$}$ solves the GI instances obtained for strongly-regular graphs from 
\cite{SpenceLib} ($n\!\le\! 64$) in a reasonable time. We substantially reduce running 
time of the algorithm for the instances 
using the following points $\varepsilon^{(i)}$: $$\varepsilon^{(i)}=\varepsilon^{(i-1)}+\varepsilon_i\biggl (e_i+\alpha\sum\limits_{j\in N(i)}e_j\biggr ),$$ 
where $0\!<\!\alpha\!<\!1$ and $N(i)$ denotes the vertices that are adjacent to $i$.
Using PC, it takes not more than few minutes to solve an instance.

\vfil

\newpage

\begin{codebox}
\Procname{
$\proc{Algorithm $2$}\ (G, H)$}
\li $J\!\leftarrow\! V(H)$;
\li \For $i\!\leftarrow\! 1$ \To $n$ \Do
\li choose at random $\varepsilon_i\!\in\! S$;
\li \If $\bigl (\exists j\!\in\! J : \eta_{\mbox{\tiny{$G$}}}(\varepsilon^{(i)})\!
=\!\eta_{\mbox{\tiny{$H$}}}(\varepsilon_{\varphi}^{(i-1)}+\varepsilon_i e_j)\bigr )$ 
\li \Then $\varepsilon^{(i)}:=\varepsilon^{(i-1)}+\varepsilon_{i}e_i$; 
\li $\varphi(i)\!\leftarrow\! j$;
\li $J\leftarrow J\!\setminus\!\{j\}$; 
\li               \Else {\bf print} ``$G\!\not\simeq\! H$''.
       \End 
    \End
\li {\bf print} ``$G\!\simeq\! H$, $\varphi$ is an isomorphism of $G$ and $H$".
\end{codebox}

\bigskip

\begin{codebox}
\Procname{
$\proc{Set the correspondence}\ (i\in V(G))$}
\li \If $i\!=\!n$
\li \Then $flag\!\leftarrow\! true$;
\li       $\varphi(n)\!\leftarrow\! k$, where $k$ such that $J=\{k\}$;
\li       {\bf exit}. 
\li \Else 
\li      \For $j\!\leftarrow\! 1$ \To $n$ \Do
\li 		       choose at random $\varepsilon_i\!\in\! S$;
\li            \If $\bigl ( j\!\in\! J$ and $\eta_{\mbox{\tiny{$G$}}}(\varepsilon^{(i)})
\!=\!\eta_{\mbox{\tiny{$H$}}}(\varepsilon_{\varphi}^{(i-1)}\!+\!\varepsilon_i e_j)\bigr )$ 
\li            \Then $\varepsilon^{(i)}:=\varepsilon^{(i-1)}+\varepsilon_{i}e_i$;
\li                  $\varphi(i)\!\leftarrow\! j$;
\li                  $J\leftarrow J\!\setminus\!\{j\}$;
\li                  $\proc{Set the correspondence}\ (i+1)$;
\li                  \If $flag\!=\!false$
\li                  \Then $J\!\leftarrow\! J\!\cup\!\{j\}$;
\li                        $\varphi(i)$ is not defined;
                     \End
               \End
           \End
    \End
\li {\bf exit}.
\end{codebox}

\smallskip

\begin{codebox}
\Procname{
	$\proc{Algorithm $3$}\ (G, H)$}
\li $J\!\leftarrow\! V(H)$;
\li $flag\!\leftarrow\! false$;
\li $\forall i\!\in\! V(G) : \varphi(i)$ is not defined.
\li $\proc{Set the correspondence}\ (1)$;
\li \If $flag$ 
\li \Then {\bf print} ``$G\!\simeq\! H$, $\varphi$ is an isomorphism of $G$ and $H$";   
\li \Else 
\li       {\bf print} ``$G\!\not\simeq\! H$".
\End
\end{codebox}

\vfil

\newpage

\subsection{The probability of mistake} 

Suppose that we have some numerical realizations of the algorithms presented above, i.e., we can check equality of graphs polynomials numerically. Let $\mbox{\sffamily P}[\ \cdot \ ]$ denote a probability of the event 
that we specify in square brackets. The following theorem \cite{Schwartz}, \cite{Zippel} is known:

\begin{theorem} Let $f\!\in\! F[x_1,\ldots, x_n]$ 
be a non-zero polynomial of total degree $d\!\ge\!0$ over a field $F$. Let $S$ be a 
finite subset of $F$ and let $\varepsilon_1,\ldots,\varepsilon_n$ 
be selected at random independently and uniformly from $S$. Then
$$
\mbox{\sffamily P}[f(\varepsilon_1,\ldots,\varepsilon_n)=0]\le\frac{d}{|S|}. 
$$
\end{theorem}

If we have set up $\varphi$ for every $i\!=\!1,\ldots,n$, then we have the equality
$$\eta_{\mbox{\tiny{$G$}}}(\varepsilon_1,\ldots,\varepsilon_n)=
\eta_{\mbox{\tiny{$H$}}}(\varepsilon_{\varphi(1)},\ldots,\varepsilon_{\varphi(n)}).$$
Let $$f(\varepsilon_1,\ldots,\varepsilon_n)\!=
\!\eta_{\mbox{\tiny{$G$}}}(\varepsilon_1,\ldots,\varepsilon_n)-
\eta_{\mbox{\tiny{$H$}}}(\varepsilon_{\varphi(1)},\ldots,\varepsilon_{\varphi(n)}).$$
Total degree $d$ of the polynomial $f$ is equal to $n$, and 
$F\!=\!\mathbb{R}$ in this case. So, implementing $\proc{Algorithm $2$}$ 
or $\proc{Algorithm $3$}$, if we obtain a message ``$G\!\simeq\!H$" 
and $\varphi$ is an isomorphism of the graphs, we have $\eta_{\mbox{\tiny{$G$}}}\!\neq\!\eta_{\mbox{\tiny{$H$}}}$ with the probability 
$\mbox{\sffamily P}[mistake]\!\le\!n/10^N.$ If we set $N\!=\!n$, then  
$\mbox{\sffamily P} [mistake]\!\le\!1/10^{n-\lg n}$ and the message is 
correct with probability no less than $1\!-\!1/10^{n-\lg n}$. 

For the Direct algorithm, the message that the graphs are not isomorphic may be incorrect. The message is always 
correct for its recursive modification since in this case there is no such 
$\varphi$ that (\ref{3.1}) holds successively for $i=1,\ldots,n$. By the Theorem 1, 
it follows that the graphs are not isomorphic.

\smallskip

Let us conclude this in the Proposition 2.

\begin{proposition} 
\begin{itemize}
\item[1.] The Direct algorithm may give a wrong conclusion that\\ $G\!\not\simeq\! H$, when $G\!\simeq\! H$.

\smallskip

\item[2.] The probability of a wrong conclusion that $G\!\simeq\!H$, when $G\!\not\simeq\! H$, is negligible for the Direct algorithm.

\smallskip

\item[3.] Recursive version of the algorithm always give correct conclusion that $G\!\not\simeq\! H$. The probability of a wrong conclusion that $G\!\simeq\!H$, when $G\!\not\simeq\! H$, is negligible for the recursive version. 
\end{itemize}

\end{proposition} 
\section{Numerical realization of the algorithm. Complexity of checking polinomials for equality at some points}

The algorithms above are impractical if we can not check the equality of graph polynomials at some points  numerically using machine numbers with restricted mantissa length. We can not do it computing the value $\det(A(G)+X)$ directly since the exact computation of determinant is of exponential time complexity.

We give an approach to check the equality of values of two graph polynomials by checking not the equality of the graphs' polynomials values itself but the equality of the ratio of values of polynomials of its vertex-deleted subgraphs to values of polynomials of the graphs. We show that, in this way, we can check the equality of values of two polynomials at some points in $O(n^4)$ time using machine numbers with $O(n^2)$ mantissa length. A mantissa of this length is required to perform computations on iterations of a numerical method for solving a system of linear algebraic equations that we perform to compare the values of the polynomials. 

Our computational approach to solve GI may be considered as a process of consistent perturbations of matrices of the two graphs that we check for isomorphism. The {\it perturbations} are changes of diagonal elements of the matrices. We call the perturbations of two graph matrices {\it consistent}, if, after their implementation, we obtain the same values of the ratio of values of polynomials of its vertex-deleted subgraphs to values of polynomials of the graphs.

In general, it may be described as follows. If, in the course of iterations of the algorithm, while setting the bijection $\varphi : V(G)\to V(H)$, we can do a series of $n$ consistent perturbations of the graphs' matrices, then the graphs are isomorphic with negligible probability of mistake. If we can not do this, without mistake, they are not isomorphic.

\subsection{Modification of adjacency matrices} 

Let $d$ be the maximum degree of vertices of $G$ and $H$: $d\!=\!\max\{d_1,\ldots, d_n\}$. We suppose that 
$G$ and $H$ have the same degree sequences. Let $A$ and $B$ be modified graph 
adjacency matrices of the following form
\begin{equation}\label{3.2}
A:=A(G)+2dE,\quad B:=A(H)+2dE.
\end{equation}
$A(H)\!=\!P_{\varphi}A(G)P_{\varphi}^\top$ for some bijection $\varphi$ from $V(G)$ to $V(G)$ if and only if $A\!=\!P_{\varphi}BP_{\varphi}^\top$. 

If $d\!>\!0$, then $A$ and $B$ have the strong diagonal predominance:
$$(A)_{ii}=2d\ge 2d_i>d_i=\sum\limits_{\stackrel{j=1}{j\neq i}}^n(A)_{ij},$$
so $$(A)_{ii}-\sum\limits_{\stackrel{j=1}{j\neq i}}^n(A)_{ij}>0,$$ $i\!=\!\overline{1,n}$.
Thus, for the graphs, their matrices of the form (\ref{3.2}) satisfy the Hadamard conditions, and 
we have $\det A\!\neq\! 0$, $\det B\!\neq\! 0$ \cite{Gantmaher}. 

So, for numerical realization of the algorithms given above, we
use modified characteristic polynomials of the following form:
\begin{equation}\label{3.3}
\eta_{\mbox{\tiny{$G$}}}(x_1,\ldots,x_n)=\det(A+X),\quad    
\eta_{\mbox{\tiny{$H$}}}(x_{\varphi(1)},\ldots,x_{\varphi(n)})=\det(B+X_{\varphi}),
\end{equation}
where $X_{\varphi}\!=\!P_{\varphi}XP_{\varphi}^{\top}$. This modification is 
equivalent to the change of variables: $x_i\!\to\! x_i+2d$. 

\subsection{Using perturbations of the matrices to solve GI\\ by checking equality of graphs polynomials}

We may consider the $\proc{Algorithm $2$}$ and the $\proc{Algorithm $3$}$ as a try to perform
series of consistent {\it perturba\-tions} of the matrices $A$ and $B$:  
\begin{equation}\label{3.4}
A^{(i)}:=A^{(i-1)}+\varepsilon_iE_i,\quad B^{(i)}:=B^{(i-1)}+\varepsilon_iE_j,
\end{equation}
where $A^{(0)}=A,\ B^{(0)}=B$, $i\!=\!1,\ldots,n$. We call the perturbations 
of the form (\ref{3.4}) {\it consistent}, if we, for every 
$i\!\in\!V(G)$, successively choose such $j\!\in\!V(H)$ that holds
\begin{equation}\label{3.5}
((A^{(i)})^{-1})_{ii}=\frac{A_{\{i\}}^{(i)}}{\det A^{(i)}}=
\frac{(B^{(i-1)}+\varepsilon_iE_j)_{\{j\}}}
{\det(B^{(i-1)}+\varepsilon_iE_j)}=((B^{(i-1)}+\varepsilon_iE_j)^{-1})_{jj}.
\end{equation}
Here $E_i$ is $n\times n$-matrix such that all of its elements are zeros except 
the $i$-th diagonal element which is equal to $1$. It follows from (\ref{3.5}) that
\begin{equation}\label{3.6}
\eta_{\mbox{\tiny{$G$}}}(\varepsilon^{(i)})=
\eta_{\mbox{\tiny{$H$}}}(\varepsilon_{\varphi}^{(i-1)}+\varepsilon_i e_j)
\end{equation}
since the equality (\ref{3.5}) is equivalent to the equality
\begin{equation}\label{3.7}
\frac{\eta_{\mbox{\tiny{$G\!\setminus\!\{i\}$}}}(\varepsilon^{(i)})}{\eta_{\mbox{\tiny{$G$}}}(\varepsilon^{(i)})}=
\frac{\eta_{\mbox{\tiny{$H\!\setminus\!\{j\}$}}}(\varepsilon_{\varphi}^{(i-1)}+\varepsilon_i e_j)}
{\eta_{\mbox{\tiny{$H$}}}(\varepsilon_{\varphi}^{(i-1)}+\varepsilon_i e_j)},
\end{equation}
and the values
$\eta_{\mbox{\tiny{$G\!\setminus\!\{i\}$}}}(\varepsilon^{(i)})$ and 
$\eta_{\mbox{\tiny{$H\!\setminus\!\{j\}$}}}(\varepsilon_{\varphi}^{(i-1)}+
\varepsilon_i e_j)$ not change when we get $\varepsilon_i\neq 0$ since they do not have $x_i$ and $x_j$ respectively as variables. So, if the equality (\ref{3.7}) holds for $\varepsilon_i\!=\!0$ and 
for some non-zero value of $\varepsilon_i$, then the equality (\ref{3.6}) 
holds too. Thus, if we can perform series of consistent perturbations for 
$i\!=\!1,\ldots,n$, then the values of the polynomials of 
the graphs are equal at the points $\varepsilon^{(i)}$ and $\varepsilon_{\varphi}^{(i)}$. 

\smallskip

The idea of using consistent perturbations of two matrices to solve GI belongs to R.T. Faizullin. It was presented in \cite{FaizullinProlubnikov}. 

\subsection{Numerical realization of the algorithms}

Thus, to implement the Direct algorithm (the $\proc{Algorithm $2$}$) numerically, having $A^{(0)}\!=\!A$, $B^{(0)}\!=\!B$, we
\begin{itemize}
\item[1)] substitute checking the condition $$\bigl (\exists j\!\in\! J\ : \ \eta_{\mbox{\tiny{$G$}}}(\varepsilon^{(i)})\!
=\!\eta_{\mbox{\tiny{$H$}}}(\varepsilon_{\varphi}^{(i-1)}+\varepsilon_i e_j)\bigr )$$ at the step 4 of the $\proc{Algorithm $2$}$ by checking the condition $$\bigl (\exists j\!\in\! J\ : \ ((A^{(i)})^{-1})_{ii}=((B^{(i-1)}+\varepsilon_iE_j)^{-1})_{jj}\bigr );$$
\item[2)] substite the step 5 of the $\proc{Algorithm $2$}$ by $$A^{(i)}\leftarrow A^{(i-1)}+\varepsilon_iE_i,\quad B^{(i)}\leftarrow B^{(i-1)}+\varepsilon_iE_j.$$
\end{itemize}

To obtain the numerical realization of the $\proc{Algorithm $3$}$, we do the same modifications of the steps 8 and 9 of the procedure $\proc{Set the correspondence}$.

\smallskip


\begin{codebox}
\Procname{
$\proc{Algorithm $2$}\ (G, H)$}
\li $A^{(0)}\!\leftarrow\!A$, $B^{(0)}\!\leftarrow\!B$;
\li $J\!\leftarrow\! V(H)$;
\li \For $i\!\leftarrow\! 1$ \To $n$ \Do
\li choose at random $\varepsilon_i\!\in\! S$;
\li \If $\exists j\!\in\! J\, : \, ((A^{(i)})^{-1})_{ii}=((B^{(i-1)}+\varepsilon_iE_j)^{-1})_{jj}$;
\li \Then $A^{(i)}\leftarrow A^{(i-1)}+\varepsilon_iE_i, B^{(i)}\leftarrow B^{(i-1)}+\varepsilon_iE_j$; 
\li $\varphi(i)\!\leftarrow\! j$;
\li $J\leftarrow J\!\setminus\!\{j\}$; 
\li               \Else {\bf print} ``$G\!\not\simeq\! H$".
       \End 
    \End
\li {\bf print} ``$G\!\simeq\! H$, $\varphi$ is an isomorphism of $G$ and $H$".
\end{codebox}


\begin{codebox}
\Procname{
$\proc{Set the correspondence}\ (i\in V(G))$}
\li \If $i\!=\!n$
\li \Then $flag\!\leftarrow\! true$;
\li       $\varphi(n)\!\leftarrow\! k$, where $k$ such that $J=\{k\}$;
\li       {\bf exit}. 
\li \Else 
\li      \For $j\!\leftarrow\! 1$ \To $n$ \Do
\li 		       choose at random $\varepsilon_i\!\in\! S$;
\li \If $\exists j\!\in\! J\, : \, ((A^{(i)})^{-1})_{ii}=((B^{(i-1)}+\varepsilon_iE_j)^{-1})_{jj}$;
\li \Then $A^{(i)}\leftarrow A^{(i-1)}+\varepsilon_iE_i, B^{(i)}\leftarrow B^{(i-1)}+\varepsilon_iE_j$; 
\li                  $\varphi(i)\!\leftarrow\! j$;
\li                  $J\leftarrow J\!\setminus\!\{j\}$;
\li                  $\proc{Set the correspondence}\ (i+1)$;
\li                  \If $flag\!=\!false$
\li                  \Then $J\!\leftarrow\! J\!\cup\!\{j\}$;
\li                        $\varphi(i)$ is not defined;
                     \End
               \End
           \End
    \End
\li {\bf exit}.
\end{codebox}

\begin{codebox}
\Procname{$\proc{Algorithm $3$}\ (G, H)$}
\li $J\!\leftarrow\! V(H)$;
\li $flag\!\leftarrow\! false$;
\li $\forall i\!\in\! V(G) : \varphi(i)$ is not defined.
\li $\proc{Set the correspondence}\ (1)$;
\li \If $flag$ 
\li \Then {\bf print} ``$G\!\simeq\! H$, $\varphi$ is an isomorphism of $G$ and $H$";   
\li \Else 
\li       {\bf print} ``$G\!\not\simeq\! H$".
\End
\end{codebox}

\subsection{Accuracy and complexity of computations 
required for numerical realization of the approach}

We obtain element $((A^{(i)})^{-1})_{ii}$ of the matrix $(A^{(i)})^{-1}$ by
solving linear system of equations of the form
\begin{equation}\label{4.1}
A^{(i)}y=e_i,
\end{equation}
where $\{e_i\}_{i=1}^n$ is a standard basis of $\mathbb{R}^n$.
$((A^{(i)})^{-1})_{ii}$ is a value of the $i$-th component of $y$. In order to solve the systems 
of linear equations, we may use such iterative methods as the Gauss-Seidel 
method ({\it the GS-method}) or another method. The Jacobi method will be preferred to effectively leverage parallel computations. These methods converge at the rate of geometric progression 
for any starting vector because matrices of the systems have the strong diagonal 
predominance.

Using the standard numeric double type, we can solve GI instances
from \cite{Foggia} and the instances that we obtain for strongly 
regular graphs from \cite{SpenceLib}. We choose $\varepsilon_i\!\in\![\delta,1)$, 
$\delta\ge 0.001$, and use $10$ iterations of the GS-method to obtain approximate solutions for all of these instances. 

Further, proving the Propositions 3, 4 and 5, we justify the numerical 
realization of the algorithm for the general case of GI.
To do this, we must:

\smallskip

1) estimate the number of iterations that is
needed to achieve accuracy that is sufficient to check the equality (\ref{3.1});

2) estimate mantissa length of the machine numbers that is 
needed to fix the difference of the real values they are represent by obtaining approximate solutions with needed precision.

\smallskip

There are two posibilities for values of the polynomials at the initial point $0\in\mathbb{R}^n$ that we use to construct the sequence of points to check isomorphism:
\begin{itemize}
\item[1)] $\eta_{\mbox{\tiny{$G$}}}(0,\ldots,0)\neq\eta_{\mbox{\tiny{$H$}}}(0,\ldots,0)$, and the graphs are not isomorphic;
\item[2)] $\eta_{\mbox{\tiny{$G$}}}(0,\ldots,0)=\eta_{\mbox{\tiny{$H$}}}(0,\ldots,0)$, and the graphs may be  isomorphic may be not.
\end{itemize}

\noindent We cannot check neither inequality $\eta_{\mbox{\tiny{$G$}}}(0,\ldots,0)\neq\eta_{\mbox{\tiny{$H$}}}(0,\ldots,0)$ nor equalty $\eta_{\mbox{\tiny{$G$}}}(0,\ldots,0)=\eta_{\mbox{\tiny{$H$}}}(0,\ldots,0)$ directly without perturbations because of the reason we mentioned above: the exact computation of determinant is of exponential time complexity. The Proposition 3 justifies the numeric realization of the algorithm in the first case, the Proposition 4 justifies it in the second case. Also, it follows from the Proposition 4, that if $G\!\simeq\!H$, then the numerical realization of the algorithm terminates with the right 
message and the probability of mistake is negligible. 

\smallskip

Let $\mbox{\sffamily P}\bigl (\mbox{\it mistake}\bigr)$ be the probability of a wrong conclusion that $G\!\!\simeq H$ for the numerical realization of the \proc{Algorithm $3$}.

\begin{proposition} 
Let $G\!\not\!\simeq H$ and $\eta_{\mbox{\tiny{$G$}}}(0,\ldots,0)\!\neq\!\eta_{\mbox{\tiny{$H$}}}(0,\ldots,0)$.
Then, if $N\!=\!n$, $$\mbox{\sffamily P}\bigl (\mbox{\it mistake}\bigr)\le\frac{1}{10^{n(n-\lg n)}}. 
$$
\end{proposition}

\begin{proof} The \proc{Algorithm $3$} sets a correspondence $\varphi$ for
$i\!=\!\overline{1,n}$ and gives the wrong message that $G\!\simeq\!H$ only if, 
for subsequent $n$ iterations of the algorithm, we have
$$\frac{\eta_{\mbox{\tiny{$G\!\setminus\!\{i\}$}}}(\varepsilon^{(i)})}{\eta_{\mbox{\tiny{$G$}}}(\varepsilon^{(i)})}=\frac{\eta_{\mbox{\tiny{$H\!\setminus\!\{\varphi(i)\}$}}}(\varepsilon_{\varphi}^{(i)})}{\eta_{\mbox{\tiny{$H$}}}(\varepsilon_{\varphi}^{(i)})},$$
and $\eta_{\mbox{\tiny{$G$}}}\!\not\equiv\!\eta_{\mbox{\tiny{$H$}}}$. 
If $\eta_{\mbox{\tiny{$G$}}}(0,\ldots,0)\!\neq\!\eta_{\mbox{\tiny{$H$}}}(0,\ldots,0)$
and  $$\frac{\eta_{\mbox{\tiny{$G\!\setminus\!\{1\}$}}}(\varepsilon^{(1)})}{\eta_{\mbox{\tiny{$G$}}}(\varepsilon^{(1)})}=\frac{\eta_{\mbox{\tiny{$H\!\setminus\!\{\varphi(1)\}$}}}(\varepsilon_{\varphi}^{(1)})}{\eta_{\mbox{\tiny{$H$}}}(\varepsilon_{\varphi}^{(1)})}$$ holds,
then, for $t=\eta_{\mbox{\tiny{$G\!\setminus\!\{1\}$}}}(\varepsilon^{(1)})/ 
\eta_{\mbox{\tiny{$H\!\setminus\!\{\varphi(1)\}$}}}(\varepsilon_{\varphi}^{(1)})$, we have
$$\eta_{\mbox{\tiny{$G$}}}(\varepsilon^{(1)})=t\cdot \eta_{\mbox{\tiny{$H$}}}(\varepsilon_{\varphi}^{(1)}).$$ Since
$\varepsilon_1$ is taken randomly from $S$, and since the values of all variables except $x_ 1$ and $x_{\varphi(1)}$ are equal to zero, then, considering $\eta_{\mbox{\tiny{$G$}}}$ and $\eta_{\mbox{\tiny{$H$}}}$ as polynomials of one variable $x_1$ and $x_{\varphi(1)}$ respectively, applying Theorem 2, we have for this case that
$$\mbox{\sffamily P}\biggl (\frac{\eta_{\mbox{\tiny{$G\!\setminus\!\{1\}$}}}(\varepsilon^{(1)})}{\eta_{\mbox{\tiny{$G$}}}(\varepsilon^{(1)})}=\frac{\eta_{\mbox{\tiny{$H\!\setminus\!\{\varphi(1)\}$}}}(\varepsilon_{\varphi}^{(1)})}{\eta_{\mbox{\tiny{$H$}}}(\varepsilon_{\varphi}^{(1)})}\biggr )=
\mbox{\sffamily P}\biggl (\eta_{\mbox{\tiny{$G$}}}(\varepsilon^{(1)})=t\cdot
\eta_{\mbox{\tiny{$H$}}}(\varepsilon_{\varphi}^{(1)})\biggr )=$$ $$=
\mbox{\sffamily P}\biggl (\eta_{\mbox{\tiny{$G$}}}(\varepsilon^{(1)})-t\cdot
\eta_{\mbox{\tiny{$H$}}}(\varepsilon_{\varphi}^{(1)})=0\biggr )
\le \frac{1}{10^N}.$$ Similarly, for $i=\overline{2,n}$,
$$\mbox{\sffamily P}\biggl (\frac{\eta_{\mbox{\tiny{$G\!\setminus\!\{i\}$}}}(\varepsilon^{(i)})}{\eta_{\mbox{\tiny{$G$}}}(\varepsilon^{(i)})}=\frac{\eta_{\mbox{\tiny{$H\!\setminus\!\{\varphi(i)\}$}}}(\varepsilon_{\varphi}^{(i)})}{\eta_{\mbox{\tiny{$H$}}}(\varepsilon_{\varphi}^{(i)})}\biggr ) \le \frac{i}{10^N}.$$
Thus, probability of the mistake may be estimated as
$$
\mbox{\sffamily P}\bigl (\mbox{\it mistake}\bigr )\le\frac{1}{10^{N}}\cdot\ldots\cdot\frac{i}{10^{N}}\cdot\ldots\cdot\frac{n}{10^{N}}.
$$ If $N\!=\!n$, then we have $$
\mbox{\sffamily P}\bigl (\mbox{\it mistake}\bigr )\le\frac{1}{10^{n(n-\lg n)}}.$$
\qed
\end{proof}

\noindent{\bf Remark.} We prove the Proposition 3 supposing that
we may check the equality $(\ref{3.7})$ numerically, i.e., that we may 
check it using machine numbers with polynomially restricted mantissa 
length.

\subsection{Computational complexity of solving the systems of linear equations with needed accuracy} 

Obtaining the values from (\ref{3.7}) as components of solution vector
of (\ref{4.1}), we make possible to estimate the number of iterations 
of the numerical method (e.g., the GS-method or the Jacobi method) 
that is needed to achieve needed accuracy of computations. The accuracy 
must be sufficient to tell the differrence between exact real values using 
machine numbers with restricted mantissa length. 

Let us estimate the computational complexity of solution of the system (\ref{4.1}) 
with needed accuracy. Let $y^{(k)}$ be an approximation 
of the exact solution $y$ of the system at $k$-th iteration of the GS-method. 

The following theorem is known \cite{Bakhvalov}:

\begin{theorem}
Let, for a linear system of equations $Ay\!=\!b$, 
the matrix $A$ is such that $$\sum_{j\neq i}|a_{ij}|\!\le\!\gamma |a_{ii}|,$$ $\gamma\!<\!1$, 
$i\!=\!\overline{1,n}$. Then $$\|y-y^{(k)}\|\le\gamma\|y-y^{(k-1)}\|,$$
where $\|x\|\!=\langle x,x\rangle$ for scalar product 
$\langle\, \cdot\, ,\, \cdot\, \rangle$ in Euclidean space $\mathbb{R}^n$.
\end{theorem}

\noindent For matrices of the form (\ref{3.2}), we have $\gamma\!\le\!1/2$. Consequently, 
$$|y_i-y_i^{(k)}|\le\|y-y^{(k)}\|\le\frac{\delta_0}{2^{k}},$$
where $\delta_0$ is initial approximation error. 

Let us consider the following problem. Let $a$, $b\!\in\!\mathbb{R}$ be 
some exact values of $n$-dimensional vector components, and let $a^{(k)}$, $b^{(k)}\!\in\!\mathbb{R}$ be 
their approximations which are obtained after $k$ iterations of the GS-method. Suppose we have
$$|a-a^{(k)}|\le\frac{\delta_0}{2^{k}},\qquad |b-b^{(k)}|\le\frac{\delta_0}{2^{k}},$$
and suppose there is such $\Delta\!>\!0$ that if $a\!\neq\!b$, 
then $|a-b|\!>\!\Delta$. We must estimate the number of iterations
we need to perform to tell the difference of $a,b\!\in\!\mathbb{R}$
using their approximations $a^{(k)}$ and $b^{(k)}$. 

If mantissa length of the machine numbers is sufficient to perform computations with needed accuracy and to fix the difference between the real values, then, having
$$|a-a^{(k)}|<\frac{\Delta}{4},\qquad |b-b^{(k)}|<\frac{\Delta}{4},$$ we
have
$$|a^{(k)}-b^{(k)}|>\frac{\Delta}{2}$$ and the value of 
$|a^{(k)}\!-\!b^{(k)}|$ is grows as $k$ grows. So we may state that $a\!\neq\!b$. 
Thus, if $|a-b|\!>\!\Delta\!>\!0$, then, to check the equality $a\!=\!b$, 
we must perform such a number of iterations $K$ that
$$\frac{\delta_0}{2^K}<\frac{\Delta}{4},$$ 
i.e., 
\begin{equation}\label{5.1}
K=O\biggl (\log \frac{1}{\Delta}\biggr ).
\end{equation}
With regard to the fact that
computational complexity of one GS-method iteration is equal to
$O(n^2)$, it takes $K\!\cdot\!O(n^2)$ elementary machine operations 
to obtain solution of the system (\ref{4.1}) with needed accuracy

\subsection{Computational complexity of checking equality\\
of the polynomials values at predefined points\\
and needed machine numbers mantissa length}

As stated above, to implement the algorithm numerically,
we must be able to check equalities of the form (\ref{3.1}), or,
we can say, we must be able to numerically check the inequalities
\begin{equation}\label{6.1}
\eta_{\mbox{\tiny{$G$}}}(\varepsilon^{(i)})\!\neq\!\eta_{\mbox{\tiny{$H$}}}(\varepsilon_{\varphi}^{(i)}).
\end{equation}
We need to do this at the time that is polynomial of $n$ using machine numbers with 
mantissa length that is restricted polynomially of $n$. The 
Proposition 4 justify such ability. 

To prove Propositions 4 and 5 we use the known the Gershgorin circle theorem:

\begin{theorem}
Every eigenvalue of a matrix $A$
lies within at least one of the discs with centres $a_{ii}$ 
and radii $\sum\limits_{j\neq i}|a_{ij}|$.
\end{theorem}

\noindent The Proposition 4, that we prove further, states that if
$\eta_{\mbox{\tiny{$G$}}}(\varepsilon^{(i)})\!=\!\eta_{\mbox{\tiny{$H$}}}(\varepsilon_{\varphi}^{(i)})$
for $i\!<\!k$ and, at $k$-th iteration, we have $\eta_{\mbox{\tiny{$G$}}}(\varepsilon^{(k)})\!\neq\!\eta_{\mbox{\tiny{$H$}}}(\varepsilon_{\varphi}^{(k)})$, this fact may be established numerically. 

To prove the Proposition 4, we need to prove the following lemma.

\begin{lemma} 
If (\ref{6.1}) holds, then $|\eta_{\mbox{\tiny{$G$}}}(\varepsilon^{(i)})-
\eta_{\mbox{\tiny{$H$}}}(\varepsilon_{\varphi}^{(i)}) |\!=\!p/10^{iN}\!\ge\!1/10^{iN}$, $p\!\in\!\mathbb{N}$.
\end{lemma}

\begin{proof} Let $A\!=\!A^{(i)}$. Then
\begin{equation}\label{4.3}
\eta_{\mbox{\tiny{$G$}}}(\varepsilon^{(i)})=\det A=\sum\limits_{\pi\in S_n} \biggl (\sigma(\pi)\prod\limits_{j=1}^na_{j\pi(j)}\biggr ),
\end{equation} where $\sigma(\pi)\!=\!1$, if permutation $\pi$ is even and 
$\sigma(\pi)\!=\!-1$ else. We have $$\prod\limits_{j=1}^na_{j\pi(j)}=\frac{p}{10^{k(\pi)N}},$$ 
where $p\!\in\!\mathbb{N}$ and $k(\pi)$ is a number of modified 
diagonal elements of $A^{(0)}$ that contained in the product for $\pi$
in (\ref{4.3}). Thus, 
$$\eta_{\mbox{\tiny{$G$}}}(\varepsilon^{(i)})=\sum\limits_{\pi\in S_n}\biggl (\sigma(\pi)\cdot\frac{p}{10^{k(\pi)N}}\biggr ).$$ Since $A$ has the strong diagonal 
predominance, we have $\eta_{\mbox{\tiny{$G$}}}(\varepsilon^{(i)})\!>\!0$,
and, consequently, $\eta_{\mbox{\tiny{$G$}}}(\varepsilon^{(i)})\!=\!p_1/10^{iN}$, 
$\eta_{\mbox{\tiny{$H$}}}(\varepsilon^{(i)})\!=\!p_2/10^{iN}$ 
for some $p_1, p_2\!\in\!\mathbb{N}$. Thus,
$$\biggr |\eta_{\mbox{\tiny{$G$}}}(\varepsilon^{(i)})-\eta_{\mbox{\tiny{$H$}}}(\varepsilon_{\varphi}^{(i)})\biggl |=\biggr |\frac{p_1}{10^{iN}}-\frac{p_2}{10^{iN}}\biggl |.$$
If $\eta_{\mbox{\tiny{$G$}}}(\varepsilon^{(i)})\!\neq\!\eta_{\mbox{\tiny{$H$}}}(\varepsilon_{\varphi}^{(i)})$, 
then $p_1\!\neq\!p_2$ and, consequently,
$|\eta_{\mbox{\tiny{$G$}}}(\varepsilon^{(i)})-\eta_{\mbox{\tiny{$H$}}}(\varepsilon_{\varphi}^{(i)})|\!=\!p/10^{iN}\!\ge\!1/10^{iN}$, $p\!\in\!\mathbb{N}$. \qed
\end{proof}

\begin{proposition}
If, for some $\varphi\!\in\!S_n$, we have $\eta_{\mbox{\tiny{$G$}}}(\varepsilon^{(i)})\!=\!\eta_{\mbox{\tiny{$H$}}}(\varepsilon_{\varphi}^{(i)})$ 
for $i\!<\!k$ and $\eta_{\mbox{\tiny{$G$}}}(\varepsilon^{(k)})\!\neq\!\eta_{\mbox{\tiny{$H$}}}(\varepsilon_{\varphi}^{(k-1)}+\varepsilon_ke_j)$ for all $j\in V(H)$, then 
\begin{equation}\label{4.4}
\frac{\eta_{\mbox{\tiny{$G\!\setminus\!\{k\}$}}}(\varepsilon^{(k)})}{\eta_{\mbox{\tiny{$G$}}}(\varepsilon^{(k)})}\neq\frac{\eta_{\mbox{\tiny{$H\!\setminus\!\{j\}$}}}(\varepsilon_{\varphi}^{(k-1)}+\varepsilon_ke_j)}{\eta_{\mbox{\tiny{$H$}}}(\varepsilon_{\varphi}^{(k-1)}+\varepsilon_ke_j)}.
\end{equation} 
We may check the inequality (\ref{4.4}) using $O(n^4)$ elementary 
machine operations and using machine numbers with mantissa 
length of $O(n^2)$.
\end{proposition}

\begin{proof} Let, for short, $\varepsilon_{\varphi}^{(k)}$ denotes $\varepsilon_{\varphi}^{(k-1)}+\varepsilon_ke_j$. Let $a\!=\!\eta_{\mbox{\tiny{$G$}}}(\varepsilon^{(k)})$, 
$a'\!=\!\eta_{\mbox{\tiny{$G\!\setminus\!\{k\}$}}}(\varepsilon^{(k)})$. 
It follows from the Lemma 1 that
$\eta_{\mbox{\tiny{$H$}}}(\varepsilon_{\varphi}^{(k)})\!=\!a\!+\!p/10^{kN}$, 
$\eta_{\mbox{\tiny{$H\!\setminus\!\{j\}$}}}(\varepsilon_{\varphi}^{(k)})\!=\!a'\!+\!q/10^{kN},$ 
$p, q\!\in\!\mathbb{N}$, $p\!\neq\!0$. Thus, $$\biggl  |\frac{\eta_{\mbox{\tiny{$G\!\setminus\!\{k\}$}}}(\varepsilon^{(k)})}{\eta_{\mbox{\tiny{$G$}}}(\varepsilon^{(k)})}-\frac{\eta_{\mbox{\tiny{$H\!\setminus\!\{j\}$}}}(\varepsilon_{\varphi}^{(k)})}{\eta_{\mbox{\tiny{$H$}}}(\varepsilon_{\varphi}^{(k)})}\biggr |=\biggl |\frac{a'}{a}-\frac{a'+q/10^{kN}}{a+p/10^{kN}}\biggr |=\frac{1}{10^{kN}}\cdot\frac{| a'p-aq |}{
\eta_{\mbox{\tiny{$G$}}}(\varepsilon^{(k)})
\eta_{\mbox{\tiny{$H$}}}(\varepsilon_{\varphi}^{(k)})}.$$
Taking into account the modifications of diagonal elements of
$(A(G))_{ii}$ and $(A(H))_{\varphi(i)\varphi(i)}$ that 
we have made for $i\!<\!k$ using values $\varepsilon_i\!<\!1$, 
it follows from the Gershgorin theorem that
$\eta_{\mbox{\tiny{$G$}}}(\varepsilon^{(k)})\!<\!(3d+1)^n$ and 
$\eta_{\mbox{\tiny{$H$}}}(\varepsilon_{\varphi}^{(k)})\!<\!(3d+1)^n$.
Since $a\!=\!p_1/10^{kN}$, $a'\!=\!p_2/10^{(k-1)N}$, where $p_1,p_2\!\in\!\mathbb{N}$,
then if $a'p\!-\!aq\!\neq\!0$, we have $|a'p\!-\!aq|\!\ge\!1/10^{kN}$. 
Consequently, 
\begin{equation}\label{4.5}
\biggl  |\frac{\eta_{\mbox{\tiny{$G\!\setminus\!\{k\}$}}}(\varepsilon^{(k)})}{\eta_{\mbox{\tiny{$G$}}}(\varepsilon^{(k)})}-\frac{\eta_{\mbox{\tiny{$H\!\setminus\!\{j\}$}}}(\varepsilon_{\varphi}^{(k)})}{\eta_{\mbox{\tiny{$H$}}}(\varepsilon_{\varphi}^{(k)})}\biggr |>\Delta=\frac{1}{10^{2kN}(3d+1)^{2n}}.
\end{equation}

Let us show that $a'p\!-\!aq\!\neq\!0$. The equality $a'p\!-\!aq\!=\!0$ is equivalent to
$a'/a\!=\!q/p$ which is equivalent to
\begin{equation}\label{4.6}
\frac{\eta_{\mbox{\tiny{$G\!\setminus\!\{k\}$}}}(\varepsilon^{(k)})}{\eta_{\mbox{\tiny{$G$}}}(\varepsilon^{(k)})}=\frac{\eta_{\mbox{\tiny{$H\!\setminus\!\{j\}$}}}(\varepsilon_{\varphi}^{(k)})-\eta_{\mbox{\tiny{$G\!\setminus\!\{k\}$}}}(\varepsilon^{(k)})}{\eta_{\mbox{\tiny{$H$}}}(\varepsilon_{\varphi}^{(k)})-\eta_{\mbox{\tiny{$G$}}}(\varepsilon^{(k)})}=\frac{q/10^{kN}}{\eta_{\mbox{\tiny{$G$}}}(\varepsilon^{(k)})-\eta_{\mbox{\tiny{$H$}}}(\varepsilon_{\varphi}^{(k)})}.
\end{equation} 
Since $$\eta_{\mbox{\tiny{$G$}}}(\varepsilon^{(k)})=\eta_{\mbox{\tiny{$G$}}}(\varepsilon^{(k-1)})+\varepsilon_k\eta_{\mbox{\tiny{$G\!\setminus\!\{k\}$}}}(\varepsilon^{(k-1)}),$$ $$\eta_{\mbox{\tiny{$H$}}}(\varepsilon^{(k)})=\eta_{\mbox{\tiny{$H$}}}(\varepsilon_{\varphi}^{(k-1)})+\varepsilon_k\eta_{\mbox{\tiny{$H\!\setminus\!\{j\}$}}}(\varepsilon_{\varphi}^{(k-1)}),$$ and $\eta_{\mbox{\tiny{$G$}}}(\varepsilon^{(k-1)})\!=\!\eta_{\mbox{\tiny{$H$}}}(\varepsilon_{\varphi}^{(k-1)})$, we have $$\eta_{\mbox{\tiny{$G$}}}(\varepsilon^{(k)})-\eta_{\mbox{\tiny{$H$}}}(\varepsilon_{\varphi}^{(k)})=\varepsilon_k\bigl (\varepsilon_k\eta_{\mbox{\tiny{$G\!\setminus\!\{k\}$}}}(\varepsilon^{(k-1)})-\eta_{\mbox{\tiny{$H\!\setminus\!\{j\}$}}}(\varepsilon_{\varphi}^{(k-1)})\bigr ).$$

Thus, (\ref{4.6}) is equivalent to $$\frac{\eta_{\mbox{\tiny{$G\!\setminus\!\{k\}$}}}(\varepsilon^{(k)})}{\eta_{\mbox{\tiny{$G$}}}(\varepsilon^{(k)})}=\frac{1}{\varepsilon_k},$$ i.e.,
$$\frac{\eta_{\mbox{\tiny{$G\!\setminus\!\{k\}$}}}(\varepsilon^{(k)})}{\eta_{\mbox{\tiny{$G$}}}(\varepsilon^{(k-1)})+\varepsilon_k\eta_{\mbox{\tiny{$G\!\setminus\!\{k\}$}}}(\varepsilon^{(k)})}=\frac{1}{\varepsilon_k}.$$ 
This is equivalent to $$\frac{\eta_{\mbox{\tiny{$G$}}}(\varepsilon^{(k-1)})}{\eta_{\mbox{\tiny{$G\!\setminus\!\{k\}$}}}(\varepsilon^{(k-1)})}=0.$$
But it is impossible, since, by definition (\ref{3.2}), 
the matrix $A$ has the strong diagonal predominance,
and we have $\eta_{\mbox{\tiny{$G$}}}(\varepsilon^{(k-1)})\!>\!0$. 
Thus, $a'p\!-\!aq\!\neq\! 0$.

\smallskip

Let us show that we may check the inequality (\ref{4.4}) using machine 
numbers with mantissa length $O(n^2)$ and that it takes $O(n^4)$ 
elementary machine operations.

It follows from (\ref{4.5}) that, to check (\ref{4.4}) numerically, in general, it takes such mantissa length $L$ 
that $$\frac{1}{10^L}\!<\!\frac{1}{10^{2kN}(3d+1))^{2n}}.$$ If we set $N\!=\!n$, 
then, since $d\!\le\!n$, $k\!\le\!n$, it is equivalent that
$$L\!>2n^2\!+\!2n\lg(3n+1).$$ Thus, having $L\!=\!O(n^2)$, we can check (\ref{6.1}) numerically. 

According to (21), we need $K\!=\!O(\log\frac{1}{\Delta})$ iterations to achieve needed accuracy of computations. Since computational complexity of one iteration of the GS-method is of $O(n^2)$, it follows from (\ref{4.5}) that it takes $$O(\log (10^{2kn}(3d\!+\!1)^{2n}))\cdot O(n^2)\!=\!O(n^4)$$ elementary machine operations to check (\ref{6.1}) numerically. \qed
\end{proof}

\section{Eliminating symmetries of graphs\\ and reducing the exhaustive search}

The proposed recursive algorithm scheme differs from the scheme of exhaustive 
search on all bijections of $V(G)$ onto $V(H)$ 
only in step $8$ of the procedure \proc{Set the cor\-res\-pondence} that it use.
Here, in addition to check whether current $j\!\in\!V(H)$ is not already 
setted up as an image for some vertex in $V(G)$ with label less than $i$, we 
check equality of the polynomials values at the points $\varepsilon^{(i)}$ and $\varepsilon_{\varphi}^{(i)}$ for possible bijections $\varphi$. The algorithm of the same form may be obtained for numerous graph invariant characteristics. For example, we may check the equality of adjacency matrices of the 
induced subgraphs on $i$ vertices for both input graphs. These subgraphs 
include the vertices for which the correspondence is already setted up, i.e., with labels less or equal than $i$, and all of the edges whose endpoints are these vertices. We check the equality 
of the subgraphs adjacency matrices after numbering vertices of the induced 
subgraph of $H$ according to correspondence $\varphi$ we have set to this iteration.

We say that the graph $G$ has more regular structure than, e.g., random graphs, if it has more symmetries. These {\it symmetries} are graph {\it automorphisms}, i.e., such bijections $\psi$ of $V(G)$ onto itself which preserves the adjacency relation between vertices $i,j\in V(G)$ mapped on $\psi(i),\psi(j)\in V(G)$. The {\it automorphism group} of a graph $G$ is a set of isomorphisms of the graphs onto itself. We denote it as $\mbox{Aut}(G)$: $$\mbox{Aut}(G)\!=\!\{\psi\!\in\! S_n\ |\ a_{ij}=a_{\psi(i)\psi(j)},\ i,j=\overline{1,n}\}.$$ The {\it orbit} $O_i$ of the vertex $i\!\in\!V(G)$ is the set of its images by automorphisms of $G$, i.e., $$O_i(G)\!=\!\{\psi(i)\ |\ \psi\!\in\! \mbox{Aut}(G)\}.$$

\smallskip

We, informally, call one graph invariant {\it weaker} than another if it is more easy to find two non-isomorphic graphs with the same values for the first one than for another. The exhaustive search on all bijections which is reduced by equlity checking of some invariant characteristics for the input graphs may be efficient for some restricted classes of graphs. Using some invariant characteristics, we may obtain some partitions of the graphs 
vertices that we may use to sagnificantly reduce the exhaustive search. But the main problem 
stays the same for every algorithm for GI of that sort: the more symmetries in the graph and the weaker the graph invariant that we use to check at step $8$ of the procedure 
$\proc{Set the correspondence}$, the more alternatives for 
setting of $\varphi$ we have in the course of its implementation.  
So it may be computationally hard to find among them such a bijection that is an isomorphism or 
to find out that there is no isomorphism for input graphs. 

Conversely, the graphs with less regular structure, such as, e.g., random graphs or trees, give 
GI instances that are easy to solve using known algorithms since they have 
automorphism groups of low cardinalities.  

The following lemma and its proof illustrates that, solving GI for the graphs $G$ and $H$, 
we may have alternative variants for setting isomorphism of the graphs only 
if we have such $i\!\in\!V(G)$ that $|O_i(G)|\!>\!1$.

\begin{lemma} Let $G\!\simeq\!H$, $i_1,i_2\!\in\!V(H)$. 
$i_1\!\in\!O_{i_2}(H)$ if and only if there exists a~vertex 
$j\!\in\! V(G)$ and such isomorphisms $\varphi_1,\varphi_2 : V(G)\!\to\! V(H)$
that $\varphi_1(j)\!=\!i_1$, $\varphi_2(j)\!=\!i_2$.
\end{lemma}

\begin{proof} Let $i_1\!\in\!O_{i_2}(H)$. 
It follows that there is $\psi\!\in\!\mbox{Aut}(G)$ such that 
$\psi(i_1)\!=\!i_2$. Let $\varphi_1$ be an isomorphism of $G$ onto $H$ and let
$j\!=\!\varphi_1^{-1}(i_1)$, $j\!\in\!V(H)$. Let $\varphi_2\!=\!\psi\circ\varphi_1$. 
We have $\varphi_2(j)\!=\!(\psi\circ\varphi_1)(j)\!=\!\psi(i_1)\!=\!i_2$ and
$\varphi_2$ is an isomorphism of $G$ onto $H$. 

\smallskip

Conversely, suppose there are such $i_1,i_2\!\in\!V(H)$, $j\!\in\!V(G)$, and isomorphisms $\varphi_1$, $\varphi_2$ that $\varphi_1(j)\!=\!i_1$, $\varphi_2(j)\!=\!i_2$. 
Then $\psi\!=\!\varphi_2\!\circ\!\varphi_1^{-1}\!\in\!\mbox{Aut}(H)$ and
$\psi(j)\!=\!i_2$, i.e., $i_1\!\in\!O_{i_2}(H)$.\qed
\end{proof}

In the course of the perturbations (\ref{3.4}), we subsequently obtain the 
graphs with less regular structure than the input graphs have. The matrices 
$A^{(i)}$ and $B^{(i)}$ in (\ref{3.4}) may be considered as adjacency matrices 
of the graphs $G^{(i)}$ and $H^{(i)}$. These graphs has weighted loops,
i.e., edges of the form $(j,j)\!\in\!E(G)$.  After the $i$-th iteration of 
the algorithm, we have $\psi(i)\!\neq\!j$ for all 
$\psi\!\in\!\mbox{Aut}(G^{(i)})$ since the $i$-th and the $j$-th diagonal 
elements of $A(G^{(i)})$ are not equal to each other for all $j\!\in\!V(G)$, $i\!\neq\!j$.
So we have $|O_i(G^{(i)})|\!=\!1$. And, accordingly to the Lemma 2, we 
have no more than one way to set the value of $\varphi(i)$ on the $i$-th 
iteration of the algorithm. Performing transformations (\ref{3.4})
and selecting unique values of the loops weights $\varepsilon_i$, we subsequently 
obtain such $G^{(i)}$ and $H^{(i)}$ that $G^{(i)}\simeq H^{(i)}$ and 
$$|O_j(G^{(i)})|=1, j\le i,\quad |\mbox{Aut}(G^{(i+1)})|\le |\mbox{Aut}(G^{(i-1)})|,$$ 
$$|O_{\varphi(j)}(H^{(i)})|=1, j\le i,\quad |\mbox{Aut}(H^{(i)})|\le |\mbox{Aut}(H^{(i-1)})|,$$ 
and finally, for some $t\!\le\!n\!-\!1$, we get graphs without symmetries in it, i.e., with trivial automorphism group: $$|O_j(G^{(t)})|=|O_{\varphi(j)}(H^{(t)})|=1, j=\overline{1,n},
\quad |\mbox{Aut}(G^{(t)})|=|\mbox{Aut}(H^{(t)})|=1.$$

To illustrate this, let us consider an example of the Direct algorithm operating. 
Let $G$ and $H$ be the input graphs shown on a picture below.
Reduction of the number of variants to set $\varphi$ is shown in Table 1.
After the $4$-th iteration, we have $|\mbox{Aut}(G^{(4)})|\!=\!1$. 

\setlength{\unitlength}{0.05cm}
\begin{picture}(200,60)

\put(70,10.5){\circle*{1}}
\put(68.5,5.5){$6$}
\put(70,50.5){\circle*{1}}
\put(68.5,51.5){$1$}
\put(50,30.5){\circle*{1}}
\put(46,29.5){$2$}
\put(60,30.5){\circle*{1}}
\put(62,29.5){$3$}
\put(80,30.5){\circle*{1}}
\put(75,29.5){$4$}
\put(90,30.5){\circle*{1}}
\put(91,29.5){$5$}

\put(70,10.5){\line(1,1){20}}
\put(70,10.5){\line(1,2){10}}
\put(70,10.5){\line(-1,2){10}}
\put(70,10.5){\line(-1,1){20}}

\put(50,30.5){\line(1,0){10}}
\put(80,30.5){\line(1,0){10}}

\put(70,50.5){\line(1,-1){20}}
\put(70,50.5){\line(1,-2){10}}
\put(70,50.5){\line(-1,-2){10}}
\put(70,50.5){\line(-1,-1){20}}

\put(140,10.5){\circle*{1}}
\put(138.5,5.5){$4$}
\put(140,50.5){\circle*{1}}
\put(138.5,51.5){$3$}
\put(120,30.5){\circle*{1}}
\put(116,29.5){$6$}
\put(130,30.5){\circle*{1}}
\put(132,29.5){$5$}
\put(150,30.5){\circle*{1}}
\put(145,29.5){$2$}
\put(160,30.5){\circle*{1}}
\put(160.5,29.5){$1$}

\put(140,10.5){\line(1,1){20}}
\put(140,10.5){\line(1,2){10}}
\put(140,10.5){\line(-1,2){10}}
\put(140,10.5){\line(-1,1){20}}

\put(120,30.5){\line(1,0){10}}
\put(150,30.5){\line(1,0){10}}

\put(140,50.5){\line(1,-1){20}}
\put(140,50.5){\line(1,-2){10}}
\put(140,50.5){\line(-1,-2){10}}
\put(140,50.5){\line(-1,-1){20}}

\end{picture}

\begin{table}[htbp]
\caption{Reduction of alternative variants of setting $\varphi$ for $G$ and $H$.}\vspace*{2mm}
\centering \small\label{tabl_1}
\begin{tabular}{|*{8}{c|}}
\hline   
& \multicolumn{6}{c|}{Variants of setting $\varphi$} & \\ 
\cline{2-7}
\raisebox{1.5ex}[0cm][0cm]{$i$}
& $1$ & $2$ & $3$ & $4$ & $5$ & $6$ &
\raisebox{1.5ex}[0cm][0cm]{$|\mbox{Aut}(G^{(i)})|$}\\
\hline   
 $0$ & $3,4$ & $1,2,5,6$ & $1,2,5,6$ & $1,2,5,6$ & $1,2,5,6$ 
 & $3,4$ & $16$\\
\hline   
 $1$ & $3$ & $1,2,5,6$ & $1,2,5,6$ & $1,2,5,6$ & $1,2,5,6$ 
 & $4$ & $8$\\
\hline   
 $2$ & $3$ & $1$ & $2$ & $5,6$ & $5,6$ 
 & $4$ & $2$\\
\hline   
 $3$ & $3$ & $1$ & $2$ & $5,6$ & $5,6$ 
 & $4$ & $2$\\
\hline   
 $4$ & $3$ & $1$ & $2$ & $5$ & $6$ 
 & $4$ & $1$\\

\hline   

\end{tabular}
\end{table}

\noindent In the Table 2, the values of $((A^{(i)})^{-1})_{jj}$ are shown. To compute these values, we perform $10$ iterations of the GS-method in order to solve
the systems of equations of the form (\ref{4.1}). The initial 
approximation that we use is $y^{(0)}\!=\!(1,\ldots,1)$. 

\begin{table}[htbp]
\caption{Computed values of
$((A^{(i)})^{-1})_{jj}$, $i\!=\!\overline{1,3}$.}\vspace*{2mm}
\centering \small\label{tabl_2}
\begin{tabular}{|*{8}{c|}}


\hline   

$i$ & $\varepsilon_i$ & $((A^{(i)})^{-1})_{11}$ & $((A^{(i)})^{-1})_{22}$ & $((A^{(i)})^{-1})_{33}$ & $((A^{(i)})^{-1})_{44}$ & $((A^{(i)})^{-1})_{55}$ & $((A^{(i)})^{-1})_{66}$\\

\hline   
$0$ & $0$ & $0.078$ & $0.094$ & $0.094$ & 
$0.094$ & $0.094$ & $0.078$\\
\hline   
$1$ & $0.861$ & $0.070$ & $0.095$ & 
$0.095$ & $0.095$ & $0.095$ & $0.078$\\
\hline   
$2$ & $0.672$ & $0.070$ & $0.087$ & $0.095$ & 
$0.095$ & $0.095$ & $0.079$\\
\hline   
$3$ & $0.372$ & $0.071$ & $0.087$ & $0.091$ & 
$0.094$ & $0.094$ & $0.079$\\
\hline   
$4$ & $0.475$ & $0.072$ & $0.087$ & $0.091$ & 
$0.089$ & $0.095$ & $0.080$\\

\hline   

\end{tabular}
\end{table}

\smallskip

Let us show that the reduction of variants of setting of $\varphi$ 
may be implemented numerically for the general case of GI. I.e., having the equality
\begin{equation}\label{5.1}
\frac{\eta_{\mbox{\tiny{$G\!\setminus\!\{i\}$}}}(\varepsilon^{(i-1)})}{\eta_{\mbox{\tiny{$G$}}}(\varepsilon^{(i-1)})}=\frac{\eta_{\mbox{\tiny{$G\!\setminus\!\{j\}$}}}(\varepsilon^{(i-1)})}{\eta_{\mbox{\tiny{$G$}}}(\varepsilon^{(i-1)})},
\end{equation} 
after the $(i\!-\!1)$-th iteration of the algorithm, we have
\begin{equation}\label{5.2} \frac{\eta_{\mbox{\tiny{$G\!\setminus\!\{i\}$}}}(\varepsilon^{(i)})}{\eta_{\mbox{\tiny{$G$}}}(\varepsilon^{(i)})}\neq\frac{\eta_{\mbox{\tiny{$G\!\setminus\!\{j\}$}}}(\varepsilon^{(i)})}{\eta_{\mbox{\tiny{$G$}}}(\varepsilon^{(i)})}
\end{equation} 
after the $i$-th iteration. And, having (\ref{5.2}), we can check it numerically in polynomial time using machine numbers
with polinomially restricted mantissa length.

\begin{proposition} 
Let (\ref{5.2}) holds. Then
\begin{equation}\label{5.3}
\biggr |\frac{\eta_{\mbox{\tiny{$G\!\setminus\!\{i\}$}}}(\varepsilon^{(i)})}{\eta_{\mbox{\tiny{$G$}}}(\varepsilon^{(i)})}-\frac{\eta_{\mbox{\tiny{$G\!\setminus\!\{j\}$}}}(\varepsilon^{(i)})}{\eta_{\mbox{\tiny{$G$}}}(\varepsilon^{(i)})}\biggl |>\frac{1}{3^n10^Nd^2}
\end{equation}
and we can check the inequality (\ref{5.2}) using 
$O(n^3\lg n)$ elementary machine operations and using machine numbers with mantissa 
length $O(n\lg n)$.
\end{proposition}

\begin{proof} Since $$\eta_{\mbox{\tiny{$G\!\setminus\!\{j\}$}}}(\varepsilon^{(i)})=\eta_{\mbox{\tiny{$G\!\setminus\!\{j\}$}}}(\varepsilon^{(i\!-\!1)})+\varepsilon_i\eta_{\mbox{\tiny{$G\!\setminus\!\{ij\}$}}}(\varepsilon^{(i\!-\!1)})),$$ we have 
$$\biggl 
|\frac{\eta_{\mbox{\tiny{$G\!\setminus\!\{i\}$}}}(\varepsilon^{(i)})}{\eta_{\mbox{\tiny{$G$}}}(\varepsilon^{(i)})}-\frac{\eta_{\mbox{\tiny{$G\!\setminus\!\{j\}$}}}(\varepsilon^{(i)})}{\eta_{\mbox{\tiny{$G$}}}(\varepsilon^{(i)})}\biggl |
=\biggr |\frac{\eta_{\mbox{\tiny{$G\!\setminus\!\{i\}$}}}(\varepsilon^{(i)})-
(\eta_{\mbox{\tiny{$G\!\setminus\!\{j\}$}}}(\varepsilon^{(i\!-\!1)})+\varepsilon_i\eta_{\mbox{\tiny{$G\!\setminus\!\{ij\}$}}}(\varepsilon^{(i\!-\!1)}))}{\eta_{\mbox{\tiny{$G$}}}(\varepsilon^{(i)})}\biggl |=$$

\begin{equation}\label{5.3}
=\varepsilon_i\cdot \frac{\eta_{\mbox{\tiny{$G\!\setminus\!\{ij\}$}}}(\varepsilon^{(i\!-\!1)}))}{\eta_{\mbox{\tiny{$G$}}}(\varepsilon^{(i)})}
\end{equation} since $\eta_{\mbox{\tiny{$G\!\setminus\!\{i\}$}}}(\varepsilon^{(i)})=\eta_{\mbox{\tiny{$G\!\setminus\!\{i\}$}}}(\varepsilon^{(i\!-\!1)})$ as it follows from (\ref{5.1}).

We have $\eta_{\mbox{\tiny{$G\!\setminus\!\{ij\}$}}}(\varepsilon^{(i\!-\!1)}))\ge d^{n-2}$
because Hadamard conditions are satisfied. On the other hand, by the Gershgorine theorem, 
for eigenvalues $\lambda_t$ of $A(G^{(i)})$, we have $d\!\le\!\lambda_t\!\le\! 3d+1,\ t\!=\!\overline{1,n}$. Consequently,
$d^n\!\le\!\eta_{\mbox{\tiny{$G$}}}(\varepsilon^{(i)})\!=\!\prod_{r=1}^n\lambda_t\!\le\!(3d+1)^n$.
Taking this into account and using (\ref{5.3}), we obtain
\begin{equation}\label{5.4}
\biggr |\frac{\eta_{\mbox{\tiny{$G\!\setminus\!\{i\}$}}}(\varepsilon^{(i)})}
{\eta_{\mbox{\tiny{$G$}}}(\varepsilon^{(i)})}-
\frac{\eta_{\mbox{\tiny{$G\!\setminus\!\{j\}$}}}(\varepsilon^{(i)})}{\eta_{\mbox{\tiny{$G$}}}(\varepsilon^{(i)})}\biggl |\ge\varepsilon_i\cdot\frac{d^{n-2}}{(3d+1)^n}>\frac{1}{10^N3^n\bigl (d+\frac{1}{3}\bigr )^n}.
\end{equation}

Let us estimate mantissa length $L$ of machine numbers that 
is sufficient to numerically check the inequality (\ref{5.2}). 
For this purpose, it is required that 
$$\frac{1}{10^L}\!<\frac{1}{10^N3^n\bigl (d+\frac{1}{3}\bigr )^n}.$$ This inequality 
is equivalent to the inequality
$$L\!>\! N\!+\! n\lg 3\!+n\,\lg \biggl (d+\frac{1}{3}\biggr ).$$ Since $d\!\le\!n$, if
$N\!=\!n$, then, using machine numbers with mantissa length
$L$ such that
$$L\!>\!n+\!n\lg3\!+n\lg n,$$ we can check
the inequality (\ref{5.2}) numerically, i.e., needed $L\!=\!O(n\lg n)$.
Since computational complexity of one iteration of the GS-method is of $O(n^2)$, it follows from (\ref{5.4}) that
it takes $O(\lg (3^n10^nd^n))\cdot O(n^2)=O(n^3\lg n)$ elementary machine 
operations to check the inequality (\ref{5.2}) numerically. \qed
\end{proof}

It follows from the Propositions 4 and 5, that, for graphs on $n$ vertices, it takes $$O(n^4)=\max\{O(n^4),O(n^3\lg n)\}$$ elementary machine operations and we need machine numbers with mantissa length of $$O(n^2)=\max\{O(n^2),O(n\lg n)\}$$ to perform an iteration of the presented algorithms. 

\smallskip

\noindent{\bf Remark.} The above estimates of the complexity of solving systems of linear equations with the required accuracy were obtained in order to demonstrate the fundamental possibility of numerical implementation of the presented approach. For the test instances of GI mentioned in the work, as well as for other problems of dimensions from $15$ to several thousands, the number of iterations of the GS- or Jacobi methods did not exceed $20$ when achieving the required accuracy of computations. That is, for all of these instances this number of iterations can be estimated by a constant rather than by the estimate (21). Then complexity of comparing the values of graph polynomials at two points is $O(n^2)$, and the complexity of one iteration of the $\proc{Algorithm $2$}$ and the $\proc{Algorithm $3$}$ is equal to $ O(n^3)$. Computational complexity of the polynomial Direct algortithm ($\proc{Algorithm $2$}$) is $O(n^4)$.

\section*{Conclusions}

We modify characteristic polynomial of a graph on $n$ vertices considering characteristic polynomial of a graph as polynomial of $n$ variables. Assigning the modified characteristic polynomials for graphs, we reduce the graph isomorphism problem to the problem of the polynomials isomorphism checking. It is required to find out, is there  such a numbering of the second polynomial's variables that the modified characteristic polynomials of the graphs are equal. We consider such numbering as an isomorphism of the polynomials. We prove that two graphs are isomorphic if and only if the graphs' polynomials are isomorphic. We present algorithms for the graph isomorphism problem that use the reduction.

Since, for a graph on $n$ vertices, the graph polynomial has $2^n$ coefficients, its value at some point cannot be evaluated directly for large enough $n$. We prove the propositions that justify the numerical realization of the presented algorithms for the graph isomorphism problem. We show that we may check the equality of the polynomials at some points without direct evaluation of the polynomials values at these points. We prove that, for the graphs on $n$ vertices, it is required $O(n^4)$ elementary machine operations and it is required machine numbers with mantissa length $O(n^2)$ to check the equality of the polynomials values numerically. 

\smallskip

For the worst, it needs an exponential from $n$ time to solve instance of the graph isomorphism problem using the presented approach, but in practice, it is efficient even for well known compuationally hard instances of the graph isomorphism problem.

%

%
\clearpage
\end{document}